# Estimation of the precision of a structured light system in oil paintings on canvas

David Sánchez-Jiménez[1], Fernando Buchón-Moragues[1], José M. Bravo[2] and Juan V. Sánchez-Pérez[2,*]

[1] Departamento de Ingeniería Cartográfica, Geodesia y Fotogrametría, Universitat Politécnica de Valéncia, Camino de Vera s/n, 46022 Valencia, Spain; dasanji@doctor.upv.es; (D.S.-J.); fbuchon@upv.es (F.B.-M.)

[2] Centro de Tecnologías Físicas, Acústica, Materiales y Astrofísica, División Acústica, Universitat Politécnica de Valéncia, Camino de Vera s/n, 46022 Valencia, Spain; jobrapla@upv.es (J.M.B.); jusanc@fis.upv.es (J.V.S.-P.)

**Abstract:** The conservation and authentication of pictorial artworks is considered an important part of the preservation of the cultural heritage. The use of non-destructive testing allows the obtaining of accurate information about the state of pictorial artworks, without direct contact between the equipment used and the sample. In particular, the use of this kind of technology is recommended in obtaining three-dimensional surface digital models, as it provides high-resolution information that constitutes a kind of fingerprint of the samples. In the case of pictorial artworks with some kind of surface relief, one of the most useful technologies is structured light (SL). In this paper the minimum difference in height that can be distinguished with this technology is estimated, establishing experimentally both the error committed in the measurement process and the precision in the use of this technology. The study, focused on the case of oil paintings on canvas, has been developed using a low-cost system to ensure its wide use.

## 1. Introduction

Pictorial artworks are a fundamental part of the cultural heritage and therefore a great effort must be made to ensure their durability, protecting them from damage caused by external agents such as temperature, humidity and different kinds of pollution. To do that, the need for preventive conservation of this kind of artworks becomes a crucial factor, with continuous monitoring or follow-up of environmental temperature and humidity. Thus, their exhibition and conservation have to take place in rooms with temperature and humidity control systems. A temperature of 19-24ºC and a humidity of 45-65% are considered ideal conditions for conservation [1]. If these conditions are not met, the deterioration of the work is quicker, with damage to both the substrate (curvatures and deformations) and the paint, for example, cracks, incisions, tears, raised or peeled paint, loss of pigments that come off, etc. [2].

On the other hand, these measures should be supported by adequate and accurate documentation and surveying actions in order to continuously check the conservation state of the pictorial artworks ensuring, at the same time, their authenticity. To obtain this documentation, several technologies usually called non-destructive testing (NDT) allow the analysis of pictorial artworks avoiding direct contact between the sample and the instruments used. Some NDT technologies provide information about the inner layers of the pictorial artworks. Among them we can mention infrared thermography [3], X-rays [4] or ultrasounds [5]. However, obtaining

3-Dimensional Surface Digital Models (3DSDM) is recommended by some authors for the preventive conservation of pictorial artworks due to the fact that a precise 3DSDM of high resolution, made up of millions of points, possesses unique geometric information with its physical characteristics [6]. Therefore, this metric document could be considered a kind of "fingerprint" of the painting [7]. The periodic evaluation of changes in the 3DSDM could help to detect the deterioration of pictorial artworks at an early stage, improving their conservation and indicating the convenience of their restoration. Some technologies used to obtain a 3DSDM of these artworks and to model their surface layer, detecting irregularities in the canvas and its support, are raking light [8], laser scanner [9] and structured light (SL) [2, 10].

SL is a photogrammetric technology that does not make any physical contact with the object to measure, and it has been used in areas as diverse as surgery [11], industry [12] and aeronautics [13]. In addition, SL has been frequently used in the documentation of cultural heritage. For instance, without pretending to be exhaustive, in the area of archaeology to recreate details of excavated surfaces and associated artifacts in two sites of the Middle Paleolithic in South-western France [14], in the scanning of sculptures such as "Minerva of Arezzo" [15], and in architecture for obtaining three-dimensional models with photo-realistic textures of specific parts in facades of historic buildings [16] as well as in wooden maquettes of ancient Nubian temples [17]. The analysis of pictorial artworks with SL can have several useful objectives such as digitization to obtain orthoimages and high resolution three-dimensional models [1]. In this field, SL has been extensively tested and offers high resolution for analyzing relief in paintings [18], or to study deformations in Leonardo da Vinci's work "Adoration of the Magi" [19]. Some authors have proposed the use of SL in authentication tasks as identifying authorship or detecting falsifications, or in cataloging works such as classifying paintings according to the relief of their surfaces, specifically in paintings on wooden panels [7]. In addition, SL is an especially simple and fast NDT data capture system, in which it is not necessary to manipulate pictorial artworks (removal of the frame, displacement of its location, etc.) to obtain their 3DSDM, with the increase of security for the protection of the painting that this implies.

However, despite the increased use of SL in the analysis of pictorial artworks, it is remarkable the difference between the many NDT available for carrying out three-dimensional modelling studies and the scarcity of works on usage guidelines and comparative data on results obtained. On the other hand, the specifications established by the manufacturers may cause confusion among users and it is necessary to collect data, under practical conditions of use, on the usefulness, performance and accuracy of these technologies [6]. In this sense, some comparative studies have been carried out on the usefulness and reliability of SL in the area of pictorial artworks in which the precision obtained depends on the size of the object and the system used varying in a wide range, from sub-millimetric to sub-centimetric accuracy [1,2,20].

The precision of SL depends on the size and the material of the surface of the object [21]. This precision may differ from the technical specifications given by the manufacturer, which were taken in controlled conditions and for geometric known-sized objects. Following this line, and taking into account the need for estimating practically what is the real precision obtained in this kind of objects, in this paper we present an experimental study to evaluate the precision of the 3DSDM of pictorial artworks obtained with SL. In addition, we evaluate the potential use of SL to detect falsifications and to track the deterioration of original pictorial artworks over time. Among the wide range of commercial SL systems, we have focused our study on low-cost specific hardware, along with free software so that the whole system can be widely used. The results obtained are valid, due to the characteristics of SL, for pictorial artworks with some kind of relief.

This article is organized as follows: In section 2 we explain the theoretical basis of SL. The specific details of the experimental setup, the characteristics of the samples and the stages of the measurement procedure are presented in section 3. In section 4 the results obtained are shown and discussed. Finally, the last section contains the concluding remarks, where the results are summarized.

## 2. Theoretical basis

SL is a NDT based on the continuous emission of several light patterns on the sample under study, which are emitted by a video projector and captured by another video or photographic camera. With this method, the precision obtained directly depends on the spatial resolution of the captured image, which can reach hundredths of a millimetre or even microns [13, 18].

In SL the 3D coordinates of each point of an image are obtained by (i) capturing the light pattern distorted by the object; (ii) the knowledge of the angles between the projection and observation systems, and (iii) the length of the optical basis connecting the nodal points or projection centres of both systems. Using absolute coding technologies (such as Gray-Code) it is possible to automatically solve the problem of correspondence. Once this is known, the spatial position of each point is measured by triangulation. The geometric principles of SL are based on the projection of a series of binary light patterns on the object to be measured, and the subsequent analysis of its deformation. In this way, a photogrammetric triangulation is produced between the projected image (called binary light pattern), the object, and the image captured by a camera (deformed light pattern). In Figure 1a a scheme of the geometry of the process can be seen.

In order to measure this deformation (see the upper part of Figure 1b), it is necessary to know the geometry between the different elements involved: camera-projector distance (called baseline or B), the angles between the main directions of both the camera and the projector with the baseline, $\alpha$ and $\beta$ respectively, and the coordinates of both the camera (Xc,Yc,Zc) and the projector centers (Xp,Yp,Zp). All these geometrical data must be determined in a calibration process of the system on reference standards, which is carried out before the measurement. The calibration is carried out by making a set of measurements on a panel already calibrated with a known matrix of points, which is called calibration panel (see the bottom part of Figure 1b). The principles of spatial resection and intersection are then applied, in which each point on the calibration panel is defined by two lines: one defined by the projector center and the point of the image of the projected pattern, and the other defined by the camera center and the point of the captured image. Considering all these parameters, a system of equations is created, taking into account (i) the coordinates of the photographic or video camera center (Xc,Yc,Zc); (ii) the coordinates of the projector center (Xp,Yp,Zp); (iii) the image coordinates of the projected points (xp, yp); (iv) the image coordinates of the deformed points ((xc,yc); (v) the focal distances of the projector and the camera (fp, fc); (vi) the spatial position of the projector on the coordinated trihedron ($\kappa p,\phi p,\omega p$); (vii), the spatial position of the camera on the coordinated trihedron ($\kappa c,\phi c,\omega c$) and (viii) the ground coordinates of each of the measured points (X,Y,Z).

The system of equations defines the geometrical configuration of the measuring system. These parameters are unknowns in the calibration phase, and data in the measurement phase [22]. The system of equations as well as the resolution of the method can be seen in the work developed by Batlle et al [23]. Note that once the measuring system has been calibrated and the position of the projector and camera are fixed, any change affecting the relative position between them will require a new calibration. Equation 1 shows, as an example, one of the relationships between the parameters defined above

$$\begin{bmatrix} X \\ Y \\ Z \end{bmatrix} = \frac{B}{fc \, \cot \alpha - x} \begin{bmatrix} x \\ y \\ fc \end{bmatrix} \qquad (1)$$

where (X, Y, Z) are the ground coordinates of the measured point; B is the basis; $\alpha$ is the main direction of the camera with the baseline; $f_c$ is the focal distance of the camera, and (x, y) are the image coordinates in the camera.

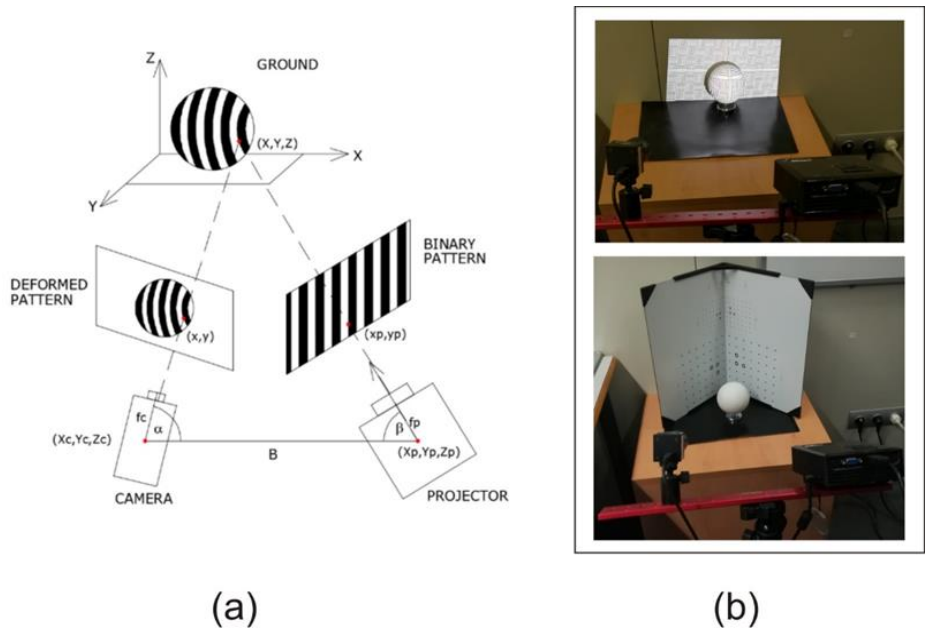

**Figure 1.** (a) Outline of the process of obtaining coordinates of points in an object using SL; (b) Two views of the experimental setup; a detail of the deformed light pattern can be seen at the upper part, while the calibration panel is shown at the bottom.

## 3. Materials and Methods

*3.1. Experimental system and dataset*

The SL scanner used in this work is the registered trademark DAVID SLS-1 [24] included in the category of active scanners based on triangulation, and the type of binary light pattern projected corresponds to the Time-Multiplexing coding [25]. This scanner is formed by a projector ACER K11+ which emits the binary patterns on the sample, and a video camera with a resolution of 1280x960 pixels with a 12 mm COMPUTAR lens that records images of the different patterns projected on the sample. To reduce the noise in the measurements, DAVID SLS-1 allows the emission of a high number of binary patterns: 58 different patterns in the high quality mode, 26 patterns in the standard mode (default option) and 22 patterns in the fast mode. The camera records each of the different projected light patterns, as well as the natural image in RGB color to assign to each measured point the color code obtained in the corresponding pixel and achieving as a result a textured 3DSDM. The maximum resolution (point density) and precision (point dispersion) achievable by DAVID SLS-1 are 0.2% and 0.1% of the object size respectively, as indicated in the manufacturer's technical specifications [26]. A great advantage of this scanner is its high precision and acceptable working speed (approximately 45 seconds for each scan) at a low price, which can be considered acceptable for its implementation in most of cataloging, restoration and authentication work.

Finally, the pieces of software used are DAVID-Laserscanner Pro Edition (DAVID Vision Systems GmbH) [24] and CloudCompare (version 2.6) [27] for obtaining and processing 3DSDM, respectively.

In this article we have worked with the high quality mode of DAVID SLS-1. Therefore, the emitter projects on each sample 58 different light patterns. This means that the resulting 3DSDM for each sample is obtained from the average of 58 measurements. Since we have worked with four samples, and each one has been measured 4 times, 16 3DSDM and 928 measurements have been obtained for the elaboration of the article. Moreover and as we will explain in the next section, we have used only four paintings but painted by the same artist, which provides the worst case situation for detecting differences.

*3.2. Characteristics of the samples*

The samples analyzed in this work are two canvases of the same size (230 mm wide x 175 mm high) stretched on a wooden support. The first canvas was measured in three successive stages simulating the different steps of the creation of a painting, in which different layers are added. These three different steps may indicate the utility of this system to track the deterioration of the painting over time. These steps are: (i) initial stage (sample A), in which a primer coat is applied; (ii) intermediate stage, in which different brushstrokes have already been applied (sample B); (iii) final stage, which corresponds to the completely finished painting (sample C). On the other hand, and to analyze the use of DAVID SLS-1 for the authentication of paintings, a second canvas was prepared imitating the first one in its final stage (sample D). Both canvases have been painted by the same artist. The four considered samples are shown in Figure 2.

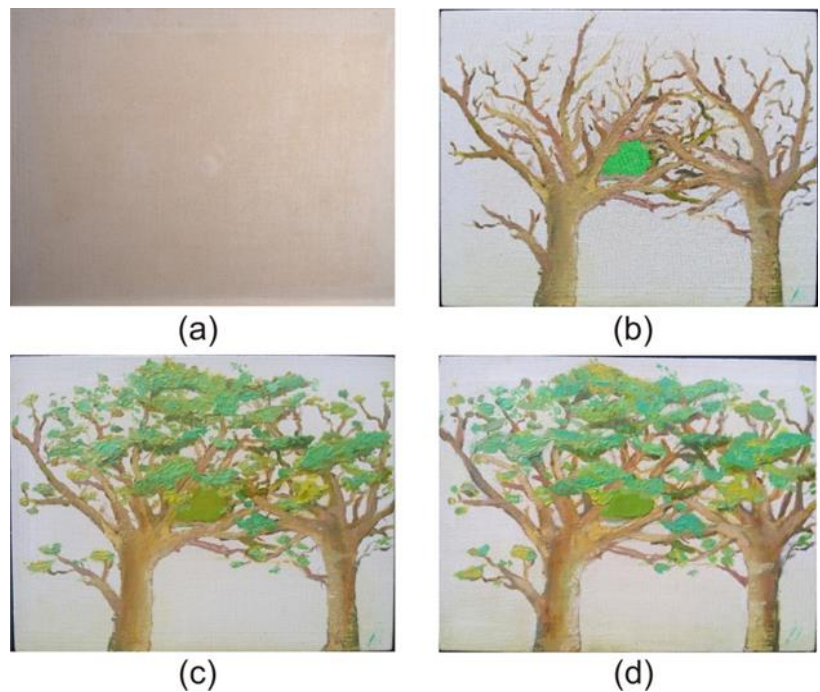

**Figure 2.** Canvases used in the study. (a) Initial state of the first canvas with a primer coat applied (sample A); (b) Intermediate state of the first canvas, in which some pictorial elements have been added (sample B); (c) Completely finished painting of the first canvas (sample C); (d) Second canvas, in which an attempt has been made to imitate the painting of the first in its final stage (sample D).

Note that, in order to control the process from the beginning, we have used the paintings created by the same artist, which is the best way to analyze the differences between two apparently equal paintings. Despite having used a relatively low dataset consisting of 16 3DSDM, they were the worst case situation, since the artist's pictorial technique does not vary and, consequently, produces more similar 3DSDM. Obviously, we could have compared a painting created by a famous artist with a copy made by our artist, but we thought that in that case the differences between the two copies would be greater, adding also the problem of the availability of the original painting. Thus, the comparison between the 3DSDM of the same two paintings created by the same artist provides the least external error for our experiment.

The maximum resolution, applying the percentages given in previous section and indicated by the manufacturer to the maximum dimensions of the canvases could be up to 0.46 mm, while the precision could be up to 0.23 mm [26].

*3.3. Methodology*

The procedure for measuring and comparing different 3DSDM using DAVID SLS-1 consists on the following steps:

(i) Calibrating the setup: Since DAVID SLS-1 does not have a fixed configuration, it needs to be calibrated each time it is assembled for measurement. To do that, the distance between the projector and the camera (baseline) must be previously determined according to the sample size. A first measurement should be made using the standard binary light patterns to obtain firstly the values of the parameters that define the photogrammetric triangulation, and secondly the coordinates of the sample in a metric coordinate system. In this work, the values of the parameters are: baseline 160 mm; distance from the scanner to the object 450 mm, calibration template 240 mm.

(ii) Obtaining the point cloud: the video projector launches a series of 58 binary light patterns with different configurations and orientations, which impinge on the sample and are deformed according to its orography. Each pattern is captured by the camera as an image and, knowing the geometry of the setup, the position of each point on the scanned object is measured, first as image coordinates, and then transformed into ground coordinates. Using DAVID-Laserscanner Pro Edition software, the point cloud is triangulated, generating a 3DSDM. Finally, the video projector launches three color patterns (blue, green and red) to obtain the texture of the object and apply it to the 3DSDM.

(iii) 3DSDM noise filtering: using CloudCompare software, continuous 3DSDM data becomes discrete, selecting 1.000.000 points evenly distributed in the sample. Point cloud noise is filtered and removed.

(iv) Comparing point clouds: two point clouds are registered: one of them is assigned as the reference cloud and the other as the comparison cloud. The aim of the registration is to minimize the distance between both point clouds, so that they locate in the same reference system and therefore become comparable.

There are several methods for registering point clouds, and many of them have been improved in recent years. An initial classification of these methods could be made according to their approach: rigid or non-rigid. Rigid methods involve a rigid environment in which a transformation of 6 degrees of freedom (DOF) takes place. It consists of a displacement and a rotation of the comparison cloud with respect to the three directions of the reference space to make it match with the reference cloud. On the other hand, non-rigid methods allow aligning objects that change their shape over time, and therefore their transformations have more than 6 DOF. Another possible classification can be made according to the required initial conditions and with the pursued detail in the register, finding rough and fine approaches. In the first one, clouds are registered roughly regardless of their initial location, while in fine approximations a more accurate registration is carried out, requiring an initial location where both clouds are close to each other. In general, both approaches are combined to reduce the number of iterations that the fine approach needs to accurately register the comparison cloud, and to increase its chances of success. In our case, we will use the method of rigid and fine approach called Iterative Closest Point (ICP) [28]. This algorithm works as follows: a cost function is defined, which represents the current error and indicates the degree of overlap between the two clouds. After that, this cost function is iteratively minimized by estimating the combination of translation and rotation that would optimally align the clouds. The least squares method is used for this purpose, in which it is possible to assign different weights at certain points and reject outliers before alignment. The translation and rotation matrix solution is obtained when several iterations are performed, or when the distance between both clouds is shorter than a certain threshold.

There are several versions of the ICP algorithm, being some of them: point-to-point ICP, which takes into account only the closest point of the comparison cloud to the reference cloud and is therefore more sensitive to outliers; point-to-surface ICP, which takes into account the vicinity of each point in both the comparison and the reference clouds, thus increasing noise resistance; non-linear ICP, which combines the solution of least squares with the sum of absolute values, and generalized ICP (G-ICP), which is more permissive in terms of ideal data assumptions and allows greater flexibility for samples with noise. Other alternatives to register point clouds are Gravitational

Approach (GA) [29], Coherent Point Drift (CPD) [30], Robust Point Matching (RPM) [31], Gaussian Mixture Model (GMM) [32], Principal Component Analysis (PCA) [33], Singular Value Decomposition (SVD) [34] and K-D tree [35].

Many comparative studies have been developed on the execution time and precision of the algorithms. Some of them were systematic reviews or benchmark surveys [36, 37] while others proposed new variants of known algorithms [38, 39]. The work of Zhu et al. [36] represents an excellent review of the available methods, in which several experiments with some representative point set registration algorithms are performed, obtaining the best results for the CPD-GL algorithm. In the study of Bellekens et al. [37], the accuracy and precision of 6 different rigid methods (PCA, SVS, 3 variants of ICP, and a combination of SVD + ICP point-to-point) is compared, with best results in accuracy for SVD + ICP point-to-point, and in precision for ICP point-to-surface. Other works present a new option to GA, with which they obtained better RMS than using other traditional methods of ICP, GA or CPD [38]. A new alternative to the SVD method has been developed by other researchers, in which point clouds are transformed into images and a SURF algorithm is used to detect homologous pixels, obtaining similar precision results to those of G-ICP, and 10 times faster [39].

As a conclusion, we have chosen the ICP algorithm for this study because it is one of the most used methods in many applications, including the 3D modelling of paintings [2, 20, 40], and it is available in CloudCompare software used in this work.

## 4. Results and discussion

In order to estimate the precision of the 3DSDM of oil paintings on canvas obtained through SL we have carried out a two-step research. The first one has been performed under controlled conditions, obtaining all the 3DSDM of the aforementioned samples using the same calibration conditions for each sample, and its objective is to estimate the experimental precision of the 3DSDM obtained with the low-cost equipment used, independently of the specifications that appear in its technical documentation. The second step has used independent calibration conditions for each 3DSDM obtained, and its goal is to check the possibility of using SL to detect falsifications or to evaluate deteriorations of pictorial artworks. This second step gives information about the experimental repeatability of the process, since it allows us to establish the percentage of similarity between two 3DSDM of the same pictorial artwork obtained on different dates. This percentage would allow us to validate if the 3DSDM obtained with SL can constitute a fingerprint of the pictorial artwork.

For that purpose, each one of the samples shown in Figure 2 has been measured four times, obtaining 16 3DSDM. The first three measurements were made without varying the position of the equipment with respect to the sample, and were used for the first experiment. These 3DSDM are named using the letter of the sample and the number of the measurement (e.g. for sample A, the first three measurements are named A1, A2 and A3). The fourth measurement of each sample was obtained after recalibrating the setup, and has been used for the second experiment. These 3DSDM measurements have been named with the letter of the sample with apostrophe (e.g. A').

To quantify the experimental precision of the SL system, we have analyzed the similarity between the measurements of a sample. To do that, we have calculated the percentage of points that have the same height in each pair of 3DSDM. This proportion is given by the cumulative relative frequency ($H_i$). This parameter is defined, for a dataset, as the number of scores that are equal to or less than a certain value. $H_i$ indicates, in the comparison of a pair of 3DSDM, the proportion of points of one of them that are at a distance shorter than a certain value in comparison with the other 3DSDM. When comparing two identical samples, the ideal result would be $H_i=1$ (i.e. 100%) for a difference in height ($\Delta h$) of 0 mm: that would mean that the point cloud of both 3DSDM is exactly in the same position. The precision of the SL system is estimated as the minimum $\Delta h$ that consistently achieves a $H_i$ close to 1 (neglecting the outliers of the 3DSDM that prevent $H_i$ from being exactly 1).

In the first step, we determine the precision of the SL system under controlled conditions; i.e., sharing calibration in their acquisition. To indirectly quantify this precision, we analyzed the $\Delta h$ between the point clouds of the different 3DSDM pairs of each sample (e. g. A1-A2; A1-A3; A2-A3), obtaining a total result of 12 combinations. The proportion of points with the same height between the different 3DSDM pairs was evaluated by means of its Hi, calculated for the Δh range between 0-0.5 mm with an interval of 0.05 mm.

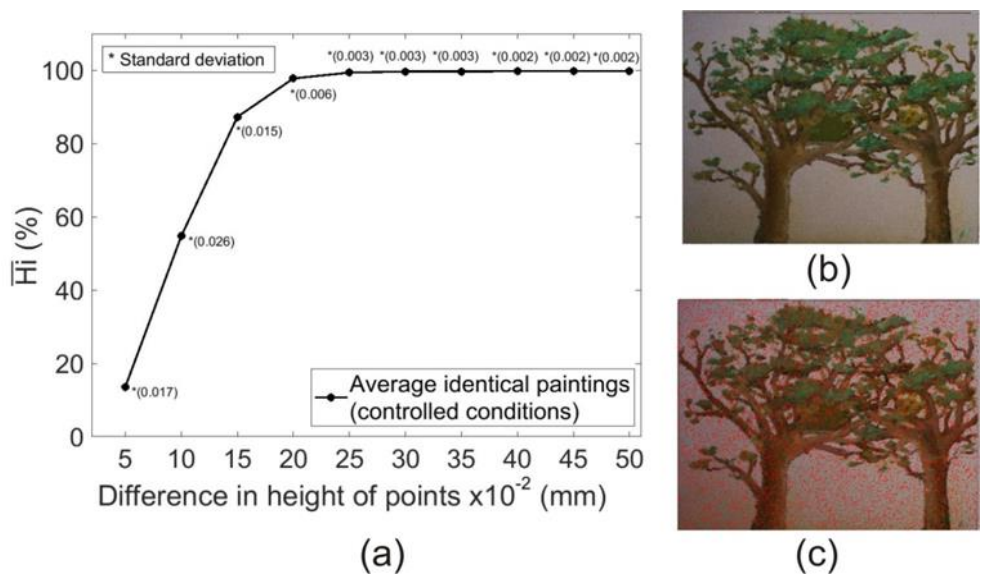

**Figure 3.** Experimental precision of the SL system analyzed with the same calibration conditions. (a) ($\overline{Hi}$) obtained for all the different 3DSDM pairs of identical samples for a Δh range of 0-0.5 mm, with an interval of 0.05 mm (n=12). (b) Example of 3DSDM comparison (Sample C: C1-C2 pair) in which red dots indicate a Δh of both longer than 0.25 mm. (c) Same 3DSDM comparison in which red dots indicate a Δh of both longer than 0.15 mm.

Figure 3a shows the arithmetic mean of Hi values ($\overline{Hi}$) obtained for each 3DSDM pair comparison of identical samples (n=12) in the Δh range of 0-0.5 mm, with an interval of 0.05 mm. One can see in the Figure that, on average, 87.3% of the points of a given 3DSDM have Δh<0.15 mm with respect to another 3DSDM of the same sample. The same applies to 99.4% of the points for Δh<0.25 mm, and in 99.8% for Δh<0.50 mm. According to these results, it can be considered reasonable to accept an experimental precision of 0.25 mm for the 3DSDM obtained with the calibration of the DAVID SLS-1 already mentioned, since the value of $\overline{Hi}$ is very close to 1 ($\overline{Hi}$=0.994 ±0.003; n=12). Note that the precision estimated by us (0.25 mm) is similar to the optimum shown in the manufacturer technical specifications, which is 0.1% of the object size (0.23 mm in our case). The low standard deviation shows the high consistency in the results obtained. Figures 3b and 3c show two representative examples of our 3DSDM pair comparison, specifically from sample C (C1 vs. C2). In these Figures, the points that show different heights comparing both 3DSDM are marked in red. It can be seen that many points show Δh>0.15 mm (Figure 3c) between both 3DSDM, but in the case of Δh>0.25 mm the number of points is residual (Figure 3b).

In a second step, we have analyzed the potential of this SL system (i.e. DAVID SLS-1) to detect falsifications and to evaluate the deterioration of pictorial artworks. This analysis gives information about two aspects. First, about the experimental repeatability of the results, comparing the 3DSDM pairs of identical samples obtained by changing the calibration of the equipment, and thus simulate that the 3DSDM have been obtained on different dates (e.g. comparing A1-A' pair). This first analysis would confirm if SL is an appropriate technology to obtain a fingerprint of a pictorial artwork. Second, by comparing the 3DSDM of different pairs of samples, it could be verified if SL is capable of detecting small differences between apparently equal samples, ensuring that this technology can be used as a tool to detect falsifications. Especially important is the comparison between the 3DSDM of the C1-D1 pair, where the differences in the relief between two equal pictorial artworks painted by the same artist would be detected. In this case, the comparison between all the samples has been taken into account (i.e. 10): 4 corresponding to the sample with different calibration conditions (A1A', B1B', C1C' and D1D'), and 6 corresponding to different samples with different calibration conditions, considering the first 3DSDM of each sample (A1B1, A1C1, A1D1, B1C1, B1D1 and C1D1).

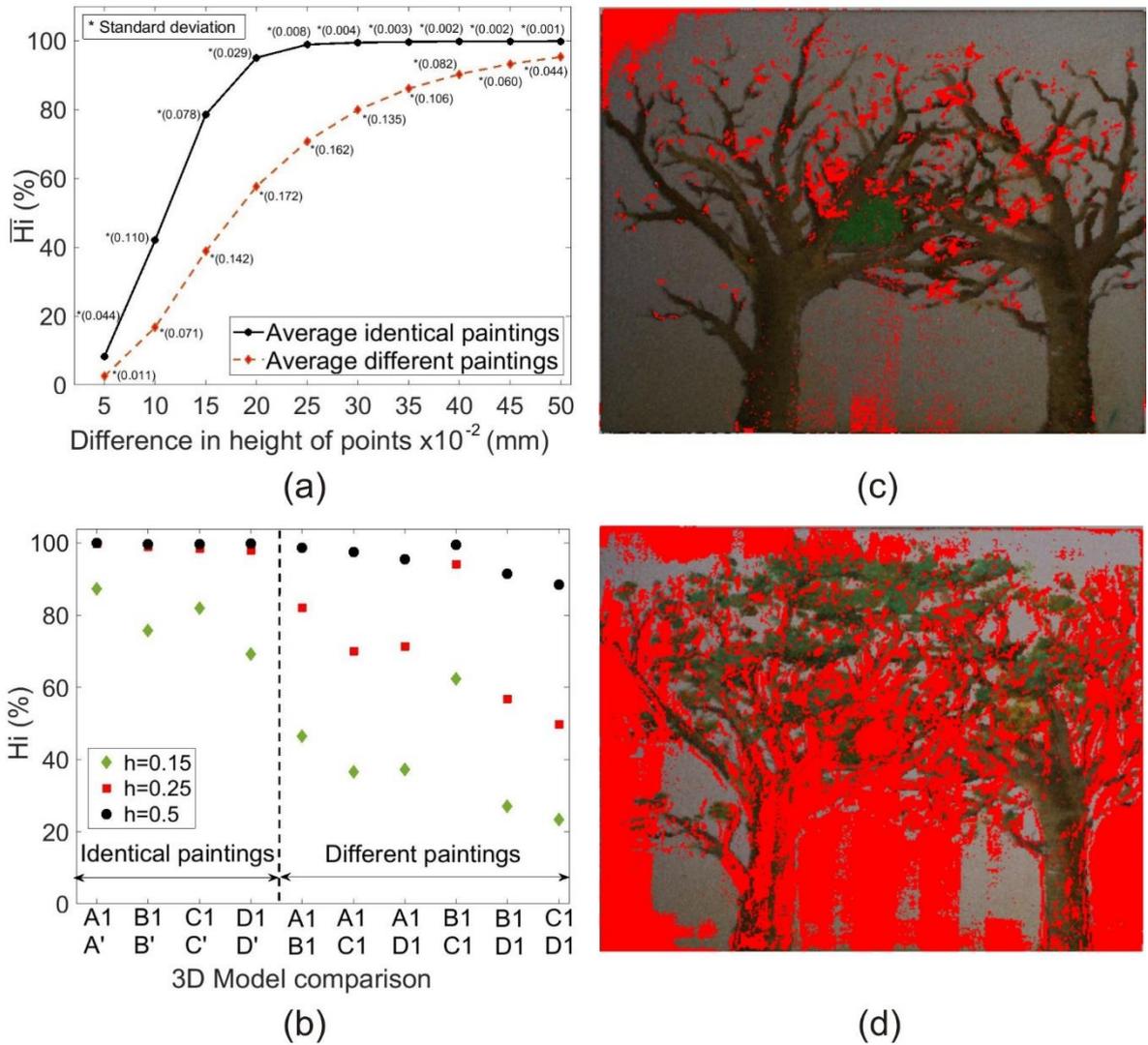

**Figure 4.** Results of the potential use of the SL system to detect falsifications or deteriorations of pictorial artworks; (a) $\overline{H_i}$ for all the different 3DSDM comparisons of identical (4 pairs, continuous black line) and different (6 pairs, discontinuous red line) samples, for a Δh range of 0-0.5 mm, with an interval of 0.05 mm (n=10). (b) 3DSDM comparison of identical (left) and different (right) samples, with indicators of their Hi for Δh=0.15 mm (green diamond), Δh=0.25 mm (red square) and Δh=0.5 mm (black circle). (c) 3DSDM comparison example between samples B and C (i.e. B1C1 pair); red dots indicate a Δh of both 3DSDM longer than 0.25 mm. (d) 3DSDM comparison example between samples C and D (i.e. C1D1 pair); red dots also indicate here a Δh of both longer than 0.25 mm.

The results of the experimental repeatability can be seen in Figure 4a where $\overline{H_i}$ has been evaluated for two cases: identical samples (n=4, e.g. A1A') and different samples (n=6, e.g. A1B1). Again, the proportion of similar points between 3DSDM of different samples has been evaluated for a Δh range of 0-0.5 mm with an interval of 0.05 mm. One can see in Figure 4a that, on average, 78.6% of the points have Δh<0.15 mm between the two 3DSDM of the same sample but with different calibration conditions. The same happens to 98.9% of the points for a Δh<0.25 mm, and in 99.9% for a Δh<0.5 mm. However, these $\overline{H_i}$ values decrease in the case of the 3DSDM comparison of different samples: 38.8% of the points for a Δh<0.15 mm, 70.8% of the points for a Δh<0.25 mm, and in 95.3% for a Δh<0.5 mm. The greatest difference between identical and different paintings is in the Δh range of 0.15-0.25 mm. However, it is necessary to point out that for Δh=0.15 mm many of the points are not identified as similar, so the sensitivity of the method would be overestimated.

On the left side of Figure 4b we show the 3DSDM comparison for pairs of identical samples with different calibration conditions, representing Hi of each pair. One can see that this value is between 60%-90% for Δh=0.15 mm, which is not acceptable for detecting falsifications or deteriorations of pictorial artworks. However, for Δh=0.25 mm, Hi values are close to the maximum possible value (Hi=1) in all the comparisons carried out, providing a high precision of the SL system even with different calibration conditions.

On the right side of Figure 4b the results of Hi values for the 3DSDM comparison of different samples with the same calibration conditions are represented. Among all the results, we would highlight B1C1. In this case, Hi value that we accepted as the precision of the method (Δh=0.25 mm) is really high. This is because the (small) differences between samples B and C are only the foliage of the trees and, consequently, the similarity is high. This result can be associated to the study of the deterioration of a pictorial artwork: small differences in the 3DSDM of the same sample obtained in different dates could mean the deterioration of some parts of the sample. The results of this 3DSDM comparison can be seen in Figure 4c, where Δh>0.25 mm are represented in red color. Some of them are due to the existence of leaves in sample C, while others indicate deteriorated areas of the painting (upper left corner and lower central part). On the other hand, while at first glance C1D1 may look similar, the analysis shows that their 3DSDM are completely different, since Δh<0.25 mm is really low (approximately 48%). This result would allow us to affirm with certainty that sample D is a bad copy of sample C. The 3DSDM comparison for this case can be seen in Figure 4d, where Δh>0.25 mm are distributed throughout the sample. Finally, note that in Figure 4a and due to the influence of different calibration conditions, $\overline{Hi}$ values for Δh=0.25 mm decrease slightly when compared to that obtained with the same calibration conditions ($\overline{Hi}$=0.994±0.003; n=12), although it is confirmed as the best to distinguish between identical ($\overline{Hi}$=0.989±0.008; n=4) and different ($\overline{Hi}$=0.708 ±0.162; n=6) samples. Other Δh are not that useful for distinguishing between identical and different samples, as Δh values shorter than 0.25 mm neglect many similar 3DSDM points of identical samples, while Δh values longer than 0.25 mm obtained more similar $\overline{Hi}$ between both groups (i.e. identical and different).

The limitations of this work include the relatively low dataset and the lack of previous reference studies of structured light systems with the aim of detection of falsifications in pictorial artworks. Because of that, the results obtained are preliminary and need to be completed with other studies that include more samples and pictorial techniques. Also, a comparison of the results obtained with the low-cost equipment used in this work with more sophisticated 3D modeling equipment, such as the high-precision triangulation laser scanner, will be considered in future studies.

## 5. Conclusions

This paper analyzes the minimum difference in height that can be distinguished with the SL system DAVID SLS-1. SL allows the obtaining of three-dimensional Surface Digital Models (3DSDM), which provide geometrical information about the surface of the pictorial artworks with high precision in an easy and economical way. This information, together with preventive conservation measures against damage caused by external agents, should constitute the measures to ensure the maintenance of the artworks over the years.

The surface information obtained with this NDT represents a fingerprint of the painting and can be used to prevent deterioration or to authenticate pictorial artworks. Obviously, determining the precision of the SL system in obtaining digital models seems essential in its use for protection purposes. In this paper, we have used the low-cost system DAVID SLS-1 to experimentally estimate its precision in oil paintings on canvas. This precision, defined as the difference in height between 3DSDM pair points, has been set by us in 0.25 mm. In other words, for differences greater than 0.25 mm, DAVID SLS-1 can be used to detect deteriorations of falsifications in this kind of pictorial artworks. Finally, we have presented an experimental example of the use of this system for authentication and conservation purposes, with good results. The methodology developed in this work can serve as a guide to determine the precision of any SL system with others more expensive and sophisticated, including new versions of DAVID-SLS.

**Author Contributions:** Conceptualization, J.V.S.-P. and J.M.B.; methodology, F.B.-M. and D.S.-J.; software, F.B.-M. and D.S.-J.; validation, F.B.-M., D.S.-J., J.V.S.-P. and J.M.B.; formal analysis, F.B.-M., D.S.-J., J.V.S.-P. and J.M.B.; investigation, F.B.-M. and D.S.-J.; data curation, F.B.-M., D.S.-J., J.V.S.-P. and J.M.B.; writing—original draft preparation, F.B.-M., D.S.-J., J.V.S.-P. and J.M.B.; writing—review and editing, F.B.-M., D.S.-J., J.V.S.-P. and J.M.B.; visualization, F.B.-M., D.S.-J., J.V.S.-P. and J.M.B.; supervision F.B.-M., D.S.-J., J.V.S.-P. and J.M.B.;

**Conflicts of Interest:** The authors declare no conflict of interest